\documentclass[aps,prb,twocolumn]{revtex4}
\usepackage{graphicx}
\usepackage{dcolumn}
\usepackage{amsmath}
\usepackage{amssymb}

\def\uu{\uparrow\uparrow}
\def\ud{\uparrow\downarrow}
\def\du{\downarrow\uparrow}
\def\dd{\downarrow\downarrow}
\def\u{\uparrow}
\def\d{\downarrow}
\def\rv{{\bf r}}
\def\rv{{\bf r}}
\def\xv{{\bf x}}
\def\kv{{\bf k}}

\begin{document}
\title{Simple Physical Picture of the
Overhauser Screened Electron-Electron Interaction}
\author{Maria Corona,$^1$ Paola Gori-Giorgi,$^1$ and John P. Perdew$^2$}
\affiliation{$^1$INFM Center for
  Statistical Mechanics and Complexity, and
Dipartimento di Fisica, Universit\`a di Roma ``La Sapienza,'' 
Piazzale A. Moro 2, 00185 Rome, Italy \\ 
$^2$Department of Physics and Quantum Theory Group, Tulane
University, New Orleans, Louisiana 70118 USA}
\date{\today}
\begin{abstract}
As shown by Overhauser and others, the pair-distribution function $g(r)$
of a many-electron system may be found by solving a two-electron
scattering problem with an effective screened electron-electron
repulsion $V(r)$. We propose a simple physical picture in which this
screened repulsion is the ``dressed-dressed'' interaction between
two neutral objects, each an electron surrounded by its full-coupling
exchange-correlation hole. For the effective interaction between
two electrons of antiparallel spin in a high-density uniform electron
gas of arbitrary spin polarization, we confirm that this picture is
qualitatively correct. In contrast, the
``bare-dressed'' interaction is too repulsive, and does not have the
expected symmetry $V_{\ud}(r) = V_{\du}(r)$.  The simple original Overhauser
model interaction, independent of the relative spin polarization $\zeta$,
does not capture the $\zeta$-dependence of the correlation contribution
to $g(r=0)$.
\end{abstract}
\maketitle
The quantum mechanical many-electron problem is notoriously hard
if all its degrees of freedom are taken into account. For both
practical computational and conceptual purposes, however, it
can often be replaced by a one- or two-electron problem with an
effective external potential or electron-electron interaction,
respectively. The effective potential that shapes the orbitals
of the one-electron problem in Kohn-Sham density functional 
theory\cite{KS,PK} has been intensively explored, but the effective
screened interaction that shapes the geminals of the two-electron
problem\cite{KO,VS,Z,RA,Ov,GS} has received less attention. Here we propose
and provide some support for a physically-appealing ``dressed-dressed''
picture based upon the interaction between two neutral objects,
each being an electron dressed by its surrounding exchange-correlation
hole. In this picture, the ``bare-bare'' Coulomb repulsion $1/r$ is
strongly screened out over the Wigner-Seitz radius $r_s$.\par

Overhauser\cite{Ov} showed that the singlet geminals of an effective
two-electron scattering problem can be used to estimate the on-top
pair-distribution function $g(0)$ in a spin-unpolarized ($\zeta=0$)
three-dimensional electron gas of uniform density
\begin{equation}
\overline{n}=3/4\pi r_s^3.
\label{eq_dens}
\end{equation}
 (We use Hartree atomic units where $\hbar=m=e^2=1$.) Overhauser
used an effective ``bare-dressed'' interaction between a bare
electron and a neutral object composed of another electron 
and a concentric sphere of positive background charge of
density $\overline{n}$ and radius $r_s$. Gori-Giorgi and
Perdew\cite{GP1,GP2} used the same effective interaction, but
solved the Overhauser model exactly and found a pair-distribution
function $g(r)$ in close agreement with that of Quantum Monte Carlo
calculations over the whole short-range region $r\lesssim r_s$,
for the physical density regime $1 \lesssim r_s \lesssim 10$. 
For the high-density
($r_s\to 0$) limit, they found good agreement with the 
exact\cite{rassolov,Geldart} 
$g(r)$ to order $r_s$. Since then, there
have been many related studies for the three- or two-dimensional
electron gas,\cite{ADPT,DPAT1,APDT,MM,nagy} sometimes using constructions
of self-consistent effective interactions following the general
``bare-dressed'' picture of Overhauser.\cite{KO,Ov} Sum rules
for the scattering phase shifts have also been derived.\cite{Z2}\par

In this work, we consider a three-dimensional uniform electron gas with
relative spin polarization 
\begin{equation}
\zeta=(\overline{n}_\u-\overline{n}_\d)/\overline{n},
\end{equation} 
where $\overline{n}=\overline{n}_\u+\overline{n}_\d$ is the total
density of Eq.~(\ref{eq_dens}). The pair-distribution function is then
\begin{equation}
g(r)=\left(\tfrac{1+\zeta}{2}\right)^2 g_{\uu}(r)
+\left(\tfrac{1-\zeta}{2}\right)^2 g_{\dd}(r)+
\tfrac{(1-\zeta^2)}{2} g_{\ud}(r),
\end{equation}
where only $g_{\ud}$ contributes at $r=0$ because of the Pauli principle.
$\overline{n}\,g(r)$ is the average density of electrons at $r$ when
an electron is at the origin, and $\overline{n}\,[g(r)-1]$ is the density
of the exchange-correlation hole at full coupling strength, which
carries a charge equal and opposite to that of the electron it 
surrounds:\cite{GP2}
\begin{equation}
\int_0^{\infty}dr\,4\pi r^2\overline{n}\,[g(r)-1] = -1,
\end{equation} 
with the same equation for $\overline{n}_\u[g_{\uu}(r)-1]$ and 
$\overline{n}_\d[g_{\dd}(r)-1]$.
We focus on the effective interaction $V_{\ud}(r)$ between two electrons
of opposite spin in the high-density ($r_s\to 0$) limit, since
in this case correlation can
be neglected and the interaction is purely electrostatic. Thus
we can evaluate the ``bare-dressed'' and ``dressed-dressed''
models exactly and compare the predictions of both to the 
exact pair-distribution function\cite{rassolov,Geldart,BPE} 
whose short-ranged part is dominated by $V_{\ud}(r)$.
We do not explicitly discuss
the electron-electron scattering effects on transport properties,
which are a second important application of the effective two-electron
problem.\cite{KO,VS,RA,nagy} We note however that the expected symmetry
$V_{\ud}=V_{\du}$ of the effective interaction for $\zeta\ne 0$
is only achieved by the ``dressed-dressed'' picture, not by the
``bare-dressed'' one.\par

{\it Spin-unpolarized gas--}
In the Overhauser approach\cite{Ov,GP1} to electronic correlation 
in the unpolarized ($\zeta=0$) uniform gas,
the many-electron problem is reduced to a scattering event between
two electrons in a suitable effective potential $V(r,r_s)$,
with a corresponding radial Schr\"odinger equation:
\begin{eqnarray}
 \left[\frac{d^2}{dr^2}-\frac{\ell(\ell+1)}{r^2}-V(r,r_s)+k^2
\right]u_{\ell} = 0, \nonumber \\
  u_{\ell}=kr\,R_{\ell}(r,k,r_s).
\label{eq_int}
\end{eqnarray} 
The presence of the other electrons is taken into account in two ways:
(i) via $V(r,r_s)$, (ii) via an average over the possible relative
momentum $k=\frac{1}{2}|\kv_1-\kv_2|$ of the scattering event. 
The exchange symmetry between
the two electrons is ensured via a proper summation over the partial
waves $\ell$; the resulting spin-resolved pair-distribution
functions are then\cite{GP1}
\begin{eqnarray}
g_{\ud}(r,r_s) & = & \left\langle\sum_{\ell=0}^{\infty}(2\ell+1) 
R_{\ell}^2(r,k,r_s)\right\rangle,
\label{eq_gudint} \\
g_{\uu}(r,r_s) & = & 2\left\langle
\sum_{\stackrel{\ell=1}{\mathrm{odd}\; \ell}}^{\infty}
(2\ell+1) R_{\ell}^2(r,k,r_s)\right\rangle,
\label{eq_guuint}
\end{eqnarray}
where the symbol $\langle\cdots\rangle$ denotes the average over
the probability $p(k)$ (obtained from the momentum distribution
of the ideal Fermi gas\cite{GP1}). Overhauser's original 
choice\cite{Ov} for $V(r,r_s)$ was the potential
of an electron surrounded by a Wigner-Seitz sphere of uniformly
distributed positive charge:
\begin{eqnarray}
V(r,r_s) = & \frac{1}{r_s}\left(\frac{1}{s}+\frac{s^2}{2}
-\frac{3}{2}\right) & (r \le r_s) \nonumber \\
& 0 & (r>r_s),
\label{eq_potOv}
\end{eqnarray}
where
\begin{equation}
s=r/r_s
\end{equation}
is a scaled variable.
As said, this simple potential gave surprisingly
accurate results\cite{GP1} for the short-range ($r\lesssim r_s$) 
part of the unpolarized-gas $g(r)$, at metallic and lower electron 
densities.  
The result for the high-density ($r_s\to 0$) limit was also
quite accurate: the form of the screened Overhauser potential ensures that
the correction to the noninteracting gas for $r_s\to 0$  is of
first order in $r_s$, as in the exact perturbative
result:\cite{rassolov} 
\begin{equation}
g_{\sigma\sigma'}(s,r_s\to 0) = g^{(0)}_{\sigma\sigma'}(s)
+r_s\,  g^{(1)}_{\sigma\sigma'}(s)+o(r_s),
\label{eq_gHD}
\end{equation}
where $g^{(0)}$ is the
pair-distribution function of the noninteracting gas. 
(Eq.~(\ref{eq_gHD}) is valid for $r\ll \sqrt{r_s}$.) In particular,
for the value of the $\ud$ pair-correlation function at contact
($r=0$), the solution of the Overhauser model gives\cite{GP1}
$g_{\ud}(r=0,r_s\to 0)=1-0.694\,r_s+o(r_s)$, in reasonable
agreement with the exact result\cite{Geldart} $1-0.732\,r_s
+o(r_s)$.
Nagy {\it et al.}\cite{nagy} have shown that the high-density
form $g_{\ud}(r=0,r_s\to 0)=1-\lambda\,r_s+o(r_s)$  
is guaranteed when Eqs.~(\ref{eq_int})-(\ref{eq_guuint}) employ a
screened potential with screening length $\propto r_s$. 
For finite $r$, the $r_s\to 0$ form of Eq.~(\ref{eq_gHD}) is satisfied,
within the Overhauser approach, if the potential $V(r,r_s)$
is such that
\begin{equation}
V(r,r_s\to 0)=\frac{1}{r_s} U(s).
\label{eq_VHD}
\end{equation}
The Overhauser potential of Eq.~(\ref{eq_potOv}) fulfills 
Eq.~(\ref{eq_VHD}) at all $r_s$.\par

{\it Spin-polarized gas --} In the original formulation of
the Overhauser model,\cite{Ov} information on
the spin polarization state of the electron gas only enters 
through the probability distribution for the relative
momentum $k$. The potential, purely based on classical electrostatic
arguments, is independent of $\zeta$. The probability
functions $p^{\sigma\sigma'}_{\zeta}(k)$  are given in
Eqs.~(42)-(44) of
Ref.~\onlinecite{GP1}, where, however,  the calculations
for the Overhauser model with $\zeta\neq 0$ have not been carried
out. Instead, a scaling relation has been proposed. 

\begin{figure}
\includegraphics[width=6.6cm]{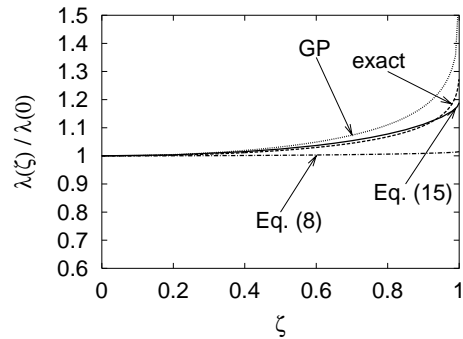} 
\caption{$\zeta$ dependence of the high-density ($r_s\to 0$) correction
to the on-top value $\lambda(\zeta)/\lambda(0)$
[see Eq.~(\ref{eq_lambda})]. 
The result from the ``dressed-dressed''
potential of Eq.~(\ref{eq_2buche}) is compared with
the exact calculation,\cite{rassolov,BPE} with the result obtained from
the original Overhauser potential of Eq.~(\ref{eq_potOv}), and with the
scaling relation proposed in Ref.~\onlinecite{GP1} (GP).}
\label{fig_lambdaz}
\end{figure}

Here, we carry out the calculations for the high-density limit
with the correct $p^{\ud}_{\zeta}(k)$, and
we find a very weak $\zeta$ dependence of the first-order correction 
$\lambda(\zeta)$ to the on-top value,
\begin{equation}
g_{\ud}(r=0,r_s\to 0,\zeta) = 1-\lambda(\zeta)\, r_s + o(r_s),
\label{eq_lambda}
\end{equation} 
as shown in Fig.~\ref{fig_lambdaz}.
This is due to the weak\cite{Ov} $k$ dependence of the short-range part
of the $s$-wave radial wavefunction $R_0(r\to 0,k,r_s)$
of Eq.~(\ref{eq_int}). An explicit dependence on $\zeta$ in the 
effective potential is thus needed in order to reproduce the
correct behavior\cite{rassolov,BPE} of the short-range 
part of $g(r)$ in the spin-polarized
electron gas. Moreno and Marinescu\cite{MM} have recently applied the
Overhauser model to the two-dimensional electron gas, finding
an extremely weak $\zeta$ dependence of the on-top value. Our
Fig.~\ref{fig_lambdaz} suggests that their result could be an artifact 
of their $\zeta$-independent effective interaction.

{\it Effective interaction for opposite-spin electrons --} 
In the high-density limit, a simple physically-motivated 
effective potential for antiparallel-spin interactions, 
which depends on $\zeta$ and has
the symmetry $\ud=\du$, can be obtained in the following way.
Consider two electrons of opposite spin in a uniform electron gas in
the high-density limit.  Each electron induces around itself an exchange
hole, forming a neutral object.  The effective potential can be
approximated with  the electrostatic interaction
between two neutral or ``dressed'' objects.
When $\zeta=0$, each electron is surrounded by a compact exchange
hole, leading to effective screening of the Coulomb repulsion.  But as
$\zeta$ approaches 1, the exchange hole around the minority spin will become
shallow and broad, so the Coulomb repulsion will be less well screened.

\begin{figure}
\includegraphics[width=6.6cm]{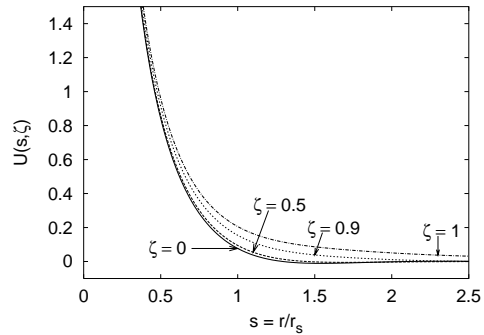} 
\caption{$\zeta$ dependence of the dimensionless
``dressed-dressed''  potential $U(s,\zeta)$
calculated from Eq.~(\ref{eq_2buche}).}
\label{fig_Uz}
\end{figure}

The two charge distributions are then
\begin{eqnarray}
\rho_1(\xv)& =& \delta(\xv)+\overline{n}_{\uparrow}[g_x^{\uu}(\xv)-1] \\
\rho_2(\xv) & = & \delta(\xv-\rv)+\overline{n}_{\downarrow}
[g_x^{\dd}(\xv-\rv)-1],
\end{eqnarray}
and the corresponding electrostatic potential is given by
\begin{equation}
V(r,r_s,\zeta) = \int d\xv\int d\xv'\, 
\frac{\rho_1(\xv)\rho_2(\xv')}{|\xv-\xv'|}.
\label{eq_2buche}
\end{equation}
$V(r,r_s,\zeta)$ can be computed analytically: its Fourier transform
$\tilde{V}(k,r_s,\zeta)$ is equal to
$$
\tilde{V}(k,r_s,\zeta)=\frac{4\pi}{k^2}+
v_1(k,r_s,\zeta)+v_2(k,r_s,\zeta)+v_3(k,r_s,\zeta),
$$
with
\begin{eqnarray}
v_1 & = & [S_x^{\uu}(k,r_s,\zeta)-1]\,\frac{4\pi}{k^2} \label{eq_1buca} \\
v_2 & = & [S_x^{\dd}(k,r_s,\zeta)-1]\,\frac{4\pi}{k^2}  \\
v_3 & = & [S_x^{\uu}(k,r_s,\zeta)-1][S_x^{\dd}(k,r_s,\zeta)-1]\,
\frac{4\pi}{k^2},
\end{eqnarray}
where $S_x^{\sigma\sigma}$ are the exchange-only static structure
factors,
\begin{eqnarray}
S_x^{\sigma\sigma}  = &  \frac{3}{4}\frac{k}{k_F^{\sigma}}
-\frac{1}{16}\left(\frac{k}{k_F^{\sigma}}\right)^3 & (k\le 2k_F^{\sigma}) 
\nonumber \\
& 1 & (k> 2k_F^{\sigma}),
\end{eqnarray}
with $k_F^{\sigma}=(1+{\rm sgn}(\sigma)\,\zeta)^{1/3}k_F$, 
$k_F=(\frac{9\pi}{4})^{1/3}r_s^{-1}$, and ${\rm sgn}(\sigma)=+1$
for spin-$\u$ and $-1$ for spin-$\d$ electrons.
The exchange-only
pair-distribution function $g_x$ only depends on $r_s$ through the
scaled variable $s=r/r_s$. This ensures that $V(r,r_s,\zeta)=
\frac{1}{r_s}U(s,\zeta)$, as required by Eq.~(\ref{eq_VHD}). 
The dimensionless potential $U(s,\zeta)$ is screened
for $s\gtrsim 1$, and goes to zero, when $s\to\infty$,
as $s^{-4}$. 
Its $\zeta$ dependence is the one
 expected from the qualitative arguments given above, as shown in
Fig.~\ref{fig_Uz}: when $\zeta\to 1$ the potential is less and less
screened; for $\zeta$ exactly equal to 1 (but only in this case)
$U(s\to\infty,\zeta=1)$ goes to zero as $s^{-2}$.

\begin{figure}
\includegraphics[width=6.6cm]{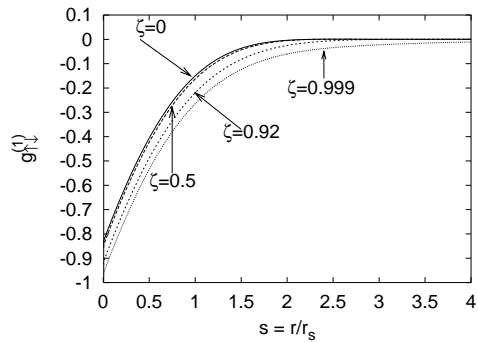} 
\caption{High-density ($r_s\to 0$) $\ud$ correlation holes computed from
the ``dressed-dressed'' potential of
Eq.~(\ref{eq_2buche}) for different values of the spin polarization $\zeta$.}
\label{fig_gcudz}
\end{figure}

Using the effective potential $U(s,\zeta)$ in the
Overhauser scheme, we calculated the 
$\ud$ high-density pair-correlation functions, $g_{\ud}^{(1)}$,
for different values of the spin-polarization $\zeta$. They are
shown in Fig.~\ref{fig_gcudz}: the qualitative behavior is very similar
to the exact one of Fig.~1 of Rassolov {\it et al.}\cite{rassolov}
This is more evident in our Fig.~\ref{fig_lambdaz}, where the 
function $\lambda(\zeta)/\lambda(0)$ is compared with the exact result.
\begin{figure}
\includegraphics[width=6.6cm]{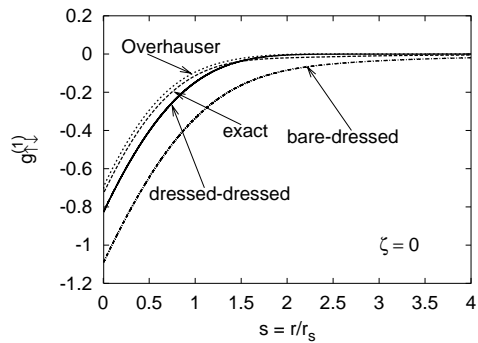} 
\caption{High-density ($r_s\to 0$) $\ud$ pair-correlation function 
for the $\zeta=0$ gas obtained from
different screened interactions:
the ``dressed-dressed'' potential
of Eq.~(\ref{eq_2buche}), the 
original Overhauser potential of Eq.~(\ref{eq_potOv}), and
the ``bare-dressed'' potential of Eq.~(\ref{eq_bd}).
The exact calculation of Rassolov et al.~\cite{rassolov} is also
reported.}
\label{fig_gcudz0}
\end{figure}
\begin{figure}
\includegraphics[width=6.6cm]{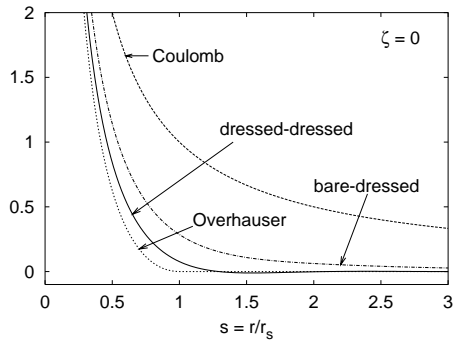} 
\caption{Comparison of the bare Coulomb potential
with different simple screened potentials for the
$\zeta=0$ gas in the high-density ($r_s\to 0$) limit: 
the ``bare-dressed'' potential of Eq.~(\ref{eq_bd}),
the ``dressed-dressed'' potential of Eq.~(\ref{eq_2buche}), and
the original Overhauser potential of Eq.~(\ref{eq_potOv}).
All curves have been multiplied by $r_s$.}
\label{fig_potentials}
\end{figure}

While the $\zeta$ dependence of $g_{\ud}^{(1)}(s,\zeta)$
obtained from the simple potential $U(s,\zeta)$ is rather good, the
quantitative agreement with the exact result when $\zeta=0$ is
less accurate than the result obtained with the original Overhauser
potential. This is shown in Fig.~\ref{fig_gcudz0}: we see that for
small $s$, $g_{\ud}^{(1)}$ obtained with $U(s,\zeta)$ 
of Eq.~(\ref{eq_2buche}) is too deep, while
the original Overhauser potential of Eq.~(\ref{eq_potOv})
gives  a result which is slightly less
deep than the exact one. This means that the original Overhauser
potential of Eq.~(\ref{eq_potOv}) is slightly too screened 
in the $r_s\to 0$ limit, while
$U(s,\zeta=0)$ of Eq.~(\ref{eq_2buche})
is not screened enough in the same limit. The ``exact''
effective potential for the high-density limit should thus lie in between
the two curves ``Overhauser'' and ``dressed-dressed'' of 
Fig.~\ref{fig_potentials}. In the same figure we also show the
bare Coulomb potential, and the ``bare-dressed'' potential
(obtained from the interaction of a ``bare'' electron with a ``dressed''
electron, i.e., surrounded by its exchange hole), whose
Fourier transform $\tilde{V}_1(k,r_s,\zeta)$ is
\begin{equation}
\tilde{V}_1(k,r_s,\zeta)=\frac{4\pi}{k^2}+v_1(k,r_s,\zeta),
\label{eq_bd}
\end{equation}
where $v_1$ is given in Eq.~(\ref{eq_1buca}).
The ``bare-dressed''
potential is ``philosophically'' closer to the original 
picture of Overhauser\cite{KO,Ov} and to
the high-density limit of the self-consistent Hartree approximation
of Davoudi {\it et al.}\cite{DPAT1}  We see that the
``bare-dressed'' potential is much
less screened that the ``dressed-dressed'' one and thus corresponds to a
a deeper (i.e., further from the exact result) $g_{\ud}^{(1)}$,
as shown in Fig.~\ref{fig_gcudz0}.\par

The ``bare-dressed'' potential encounters severe problems for the
calculation of $\lambda(\zeta)/\lambda(0)$ of Fig.~\ref{fig_lambdaz}.  
When $\zeta \to 1$, each
majority $\u$ electron dresses itself in an exchange hole deeper and more
short-ranged than for $\zeta = 0$, while each minority $\d$ electron 
undresses. So the interaction between a bare $\d$ and a 
dressed $\u$ becomes {\em less} repulsive
as $\zeta$ increases from 0, reducing $\lambda(\zeta)/\lambda(0)$.  
If we try to symmetrize using the interaction of a hypothetical 
bare $\u$ with a dressed $\d$, we find that this interaction tends 
to the unscreened $1/r$ as $\zeta \to 1$.\par

{\it Conclusions --} We have proposed a simple ``dressed-dressed'' 
picture for the
effective screened electron-electron interaction that shapes the geminals
and thus the pair distribution function of a many-electron system.  In
this picture, the interaction is between two neutral objects, each an
electron dressed by its exchange-correlation hole.  For two electrons of
opposite spin in a high-density electron gas of arbitrary spin
polarization, where the ``dressed-dressed'' and ``bare-dressed'' interactions
can be evaluated exactly, we have shown that the ``dressed-dressed'' picture
is qualitatively correct. In
future work, it may be possible to construct the ``dressed-dressed'' 
$V_{\sigma\sigma'}(r)$ for all $r_s$ and $\zeta$, using density 
functional theory\cite{KS,PK} to describe the additional 
exchange-correlation terms that arise when
$\sigma' = \sigma$ or $r_s \gg 0$.\par

MC and PG-G acknowledge discussions with G.B. Bachelet and
S. Caprara, and 
financial support from MIUR through COFIN2001.
JPP acknowledges support from the National Science
Foundation under grant DMR 01-35678.

\end{document}